\pdfoutput=1

\documentclass[prb,aps,twocolumn,amsmath,amssymb,floatfix,superscriptaddress]{revtex4}

\usepackage{color}
\usepackage{soul}
\usepackage{amsmath}
\usepackage{latexsym}
\usepackage{amssymb}
\usepackage{mathrsfs}
\usepackage{graphics,epstopdf}
\usepackage[colorlinks=true, citecolor=blue, urlcolor=blue ]{hyperref}
\usepackage{epsf,graphics,graphicx}

\date{\today}

\begin{document}
\title{Enhancement of persistent current in a non-Hermitian disordered ring}

\author{Suparna Sarkar}
\thanks{These authors contributed equally to this work.}
\author{Soumya Satpathi}
\thanks{These authors contributed equally to this work.}
\author{Swapan K. Pati}
\email{pati@jncasr.ac.in}
\affiliation
{Theoretical Sciences Unit, School of Advanced Materials (SAMat), Jawaharlal Nehru Centre for Advanced Scientific Research, Bangalore 560064, India}
\date{\today} 

\begin{abstract}
We study the Aharonov-Bohm flux-induced magnetic response of a disordered non-Hermitian ring. The disorder is
introduced through an on-site quasiperiodic potential described by the Aubry-Andr\'{e}-Harper (AAH) model, 
incorporating a complex phase that renders the model non-Hermitian. Our findings reveal that this form of 
non-Hermiticity enhances the persistent current without requiring hopping dimerization. We explore both 
non-interacting and interacting scenarios. In the former, we examine spinless fermions within the tight-binding 
(TB) framework, while in the latter, we consider spinful fermions with Hubbard interactions. We employ density-matrix 
renormalization group (DMRG) method to find the ground state energy of the Hubbard Hamiltonian for larger system 
sizes. The Non-Hermitian phase induces both real and imaginary components of the current. We thoroughly analyze the 
energy eigenspectrum, ground state energy, and persistent current in both real and imaginary spaces for various 
system parameters. Our primary goal is to investigate the combined effects of non-Hermiticity and disorder strength 
on persistent currents. We find an enhancement in both the real and imaginary components of the persistent current 
with increasing disorder strength and non-Hermiticity, up to a critical value. Furthermore, we observe an enhancement 
in the persistent current in the presence of Hubbard correlations. Our findings may provide a new route to get 
nontrivial characteristics in persistent currents for a special type of non-Hermitian systems.
\end{abstract}

\maketitle

\section{Introduction}

In the presence of magnetic flux, a normal metal ring sustains a non-dissipative current at very low temperature. 
This is known as persistent current~\cite{pc1,kulik ,levy} and it arises from the Aharonov-Bohm (AB) effect, where 
electron wave functions acquire a phase shift due to the enclosed magnetic flux. It was initially proposed by 
B\"{u}ttiker et al. in 1983~\cite{pc1} and later the first experimental confirmation came in 1990 when L\'{e}vy 
et al\cite{levy} observed persistent currents in a small metallic ring. After that, the concept of persistent current 
has inspired extensive exploration, both theoretically~\cite{pcth1,pcth2,pcth3,pcth4,pcth5,pcth6} as well as 
experimentally~\cite{pcexp1,pcexp2,pcexp3,pcexp4}, across a large variety of quantum systems. Since the response of 
persistent current (PC) with magnetic flux is very sensitive to disorder, a large number of studies have already 
explored PC in uncorrelated as well as correlated disordered rings~\cite{dis1,dis2,dis3}. Among various correlated 
systems, one of the most well-known and remarkable one is the Aubry-Andr\'{e}-Harper (AAH) model~\cite{aah1,aah2,aah3,aah4}.
It is quite obvious that the average persistent PC amplitude decreases with increasing disorder strength
~\cite{currdec1,currdec2}. Interestingly, recent studies have revealed that along with AAH disorder within 
Su-Schrieffer-Heeger (SSH) rings~\cite{ssh1} an enhancement in the amplitude of persistent currents can be obtained, 
suggesting that the interplay between hopping dimerization and quasiperiodic disorder can counteract disorder-induced 
suppression.

\par
In the studies mentioned above, all theoretical investigations are based on Hermitian models. However, very recently, 
the interplay between disorder and non-Hermiticity in systems governed by non-Hermitian Hamiltonians has attracted 
considerable attention~\cite{nh1,nh2,nh3,nh4,nh5,nh6,nh7,nh8}. Non-Hermitian Hamiltonians provide a 
powerful extension to the traditional quantum mechanics, enabling the modeling of open quantum systems
~\cite{open1,open2,open3} that interact with external environments and they are becoming popular for the wide 
array of intriguing physical phenomena that remain unattainable in their Hermitian counterparts~\cite{nh9,nh10,nh11}. 
In particular, systems exhibiting parity-time symmetry have attracted significant interest. This symmetry ensures a 
completely real energy spectrum below a critical threshold~\cite{ptsym1,ptsym2,ptsym3}. Previous studies 
have revealed unique topological features in non-Hermitian systems, such as the appearance of exceptional points under 
periodic boundary conditions and the non-Hermitian skin effect~\cite{se1,se2} under open boundary conditions. This 
field is growing rapidly because the sciences that it brings out not only enhances our theoretical insights but also drives innovation in various technological domains as 
non-Hermitian Hamiltonians are widely encountered across various physical systems, including exciton-polaritons
~\cite{expol1,expol2}, photonics~\cite{phot1,phot2}, mechanical systems~\cite{mech}, electrical circuits~\cite{elec} 
and biological networks~\cite{bio1,bio2}. Recently PC has been explored in a dimerized non-Hermitian ring
~\cite{pcnh1,pcnh2,pcnh3} where the non-Hermitian effects has been introduced either from non-reciprocal hopping terms 
or by physical gain and loss at different sites. Furthermore, the impact of non-Hermitian effects on 
persistent currents in interacting systems with nonzero electron-electron correlations remains quite underexplored. However, 
Zhang \textit{et al.}~\cite{intpc} investigated the effect of Hubbard interaction on persistent current in a Hatano-Nelson 
model. The study explored the variation of persistent current with interaction strength for different system sizes and 
found that the PC saturates for large system sizes and shows a sudden drop at the transition point which becomes sharper 
with increasing system sizes. This contrasts with the persistent current in Hermitian systems, which depends on a finite 
flux and disappears in the thermodynamic limit. Previous studies for Hermitian systems have shown that moderate interactions 
can enhance persistent currents in half-filled or partially filled ring systems, particularly in the presence of weak 
disorder~\cite{intpc1,intpc2}. While electron-electron interactions have been extensively studied in Hermitian systems
~\cite{intpc3,intpc4,intpc5}, the effects of non-Hermitian quasi periodicity in interacting systems remain largely 
unexplored.
\par

The above studies led us to explore other non-Hermitian Hamiltonians that exhibit enhanced persistent currents, even 
in the absence of dimerized hopping. To do this, we have introduced diagonal AAH disorder with complex phase factor in a one-dimensional mesoscopic ring. Recently, researchers have explored the localization 
properties in systems with this type of non-Hermitian AAH potential, focusing on how the non-Hermiticity influences edge 
states, parity-time symmetry breaking, topological phases and phase transitions, and the localization behavior of 
eigenstates~\cite{locnhaah1,locnhaah2,locnhaah3,locnhaah4,locnhaah5}. However, the impact of the imaginary phase in 
the AAH potential on the persistent current has not yet been addressed. In this work, first time we have explored the 
characteristics of persistent current by considering this form of non-Hermiticity. Moreover, the lack of research on the 
effect of non-Hermiticity on persistent current within the interacting picture presents a promising avenue for further 
investigation, as the interplay between non-Hermitian effects, quasi periodicity, and electron-electron interactions may 
lead to novel phenomena and rich physical insights.
\begin{figure}[ht]
	{\centering\resizebox*{6.5cm}{4.5cm}{\includegraphics{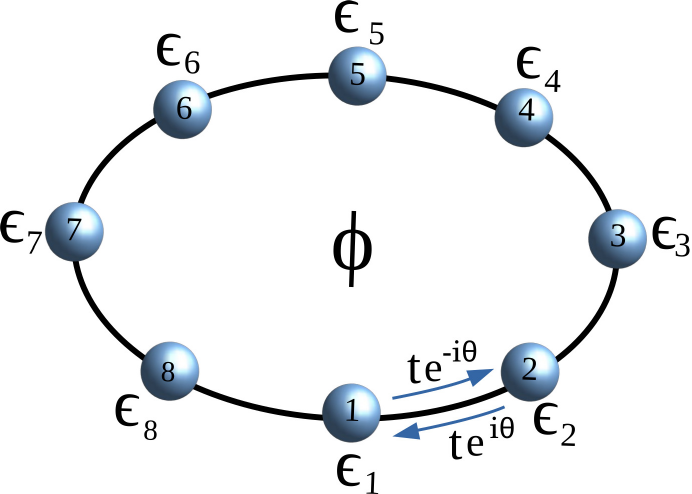}}}
	\caption{(Color online) Schematic diagram of an isolated one dimensional ring threaded by a AB flux. The ring is subjected to a quasiperiodic on-site potential containing a complex phase which makes
     the system non-Hermitian. The flux $\phi$, introduces a phase factor $\theta$ in the hopping term due to which a
     persistent current is established in the ring.}
	\label{model}
\end{figure}
In the non-interacting case, we consider spinless fermions and analyze the quantum system using a tight-binding (TB) 
framework, which provides a clear and intuitive description. The ground state energy is determined by diagonalizing 
the Hamiltonian matrix of the ring. On the other hand, for the interacting case, we introduce on-site Hubbard interactions 
between spinful fermions and describe the system by Hubbard Hamiltonian. For larger system sizes, the ground state energy 
is computed using the Density Matrix Renormalization Group (DMRG) method. Then, the PC for both the non-interacting and 
interacting scenarios, is obtained by calculating the derivative of the ground state energy with respect to AB flux. Due 
to the imaginary phase factor of the AAH potential, the PC exhibits both real and imaginary components. The imaginary component of the current represents non-conservative or non-equilibrium processes related to gain and loss 
in the system. Our study provides a comprehensive analysis of the interplay between disorder strength, non-Hermiticity, 
and electron-electron correlation. Here, we address the following important issues: (i) the dependence of ground state 
energy and persistent current on phase for different strengths of non-Hermiticity, revealing an enhancement of both the 
real and imaginary components of PC with increasing non-Hermiticity, (ii) up to a certain threshold value, both the real 
and imaginary currents increase with disorder strength in presence of non-Hermiticity, (iii) for the interacting scenario 
for a small size ring, we obtain more current for higher disorder value, and (iv) for larger system size, the DMRG results 
shows an enhancement of current with increasing disorder strength.

The structure of the remaining sections is as follows: In Section II, we describe the model, the TB Hamiltonian, the many-body 
Hamiltonian and the theoretical approach used to obtain the results. Sec. III presents all the results in an organized manner 
and examines them thoroughly. Finally, we conclude our work in Section IV.

\section{Model and Methods} \label{sec:develop}
As illustrated in Fig.~\ref{model}, we consider a non-Hermitian Aubry-Andr\'{e}-Harper (AAH) quasiperiodic ring characterized by 
complex potentials subjected to an AB flux. The ring that we have considered consists of $N$ number of lattice sites with the 
circumference $L= aN$, where $a$ is the lattice constant. The value of the lattice constant is of the order of $\mbox{\AA}$, 
so that it is small compared to other characteristic length scales of the system. This ensures that  the lattice is fine 
enough to approximate a continuous system, and the results are not distorted by the discrete nature of the model.  For the 
simplicity of theoretical calculations, in natural units, we consider the lattice constant $a = 1$, in our calculations. 
The tight-binding Hamiltonian of the ring for spinless non interacting fermions can be written as
\begin{equation}\label{eq1}
H = t\sum_{j=1}^N (e^{i\theta} c_{j}^\dagger c_{j+1}  + h.c. ) + \sum_{j=1}^N \epsilon_j n_j,
\end{equation}
where $N$ represents the total number of sites in the ring. $j$ corresponds to the lattice site index. The operators $c_{j}^\dagger (c_{j+1})$ represent the fermionic creation (annihilation) operator at site $j$ ($j+1$). Here, $t$ 
represents the hopping parameter. The AB flux $\phi$ introduces a phase factor $\theta$ in the hopping terms
which is given by $\theta=2\pi \phi/N\phi_0$. The flux $\phi$ is measured in the unit of elementary flux-quantum $\phi_0=h/e$, 
where $h$ is the Planck constant and $e$ is the electron charge. In natural units, where $\hbar=e=1$, the flux quantum simplifies 
to  $\phi_0=2\pi$ and  the phase factor becomes $\theta= \phi/N$.  $\epsilon_j$ is the on-site energy at site $j$ and 
$n_{j} = c_{j}^\dagger c_{j}$ denotes the fermionic number operator at that site. The AAH model is characterized by the functional 
form $ \epsilon_j = W \cos(2\pi b j + \phi_\nu)$, where $W$ determines the strength of the cosine modulation at 
each site. The parameter $b$ is an irrational number ($b=(\sqrt{5}-1)/2$), introducing an aperiodic nature to the lattice. The 
phase factor $\phi_\nu$, which is linked to the AAH modulation, plays a crucial role as it can be externally adjusted through an 
appropriate experimental setup \cite{pcnh2,aahph1,aahph2,aahph3}. In this context, we set $\phi_\nu = i \delta$, 
here $i$ corresponds to $\sqrt{-1}$ and the complex phase factor, $\delta$ controls the degree of non-Hermiticity within the system.

In a spinless system, following the Pauli exclusion principle, electrons sequentially occupy energy levels starting from 
the lowest available level at absolute zero temperature. The ground state energy of the system is determined by summing the 
energies of all occupied levels up to  where corresponds to the highest occupied energy level for a given number of electrons, 
which is given by
\begin{equation}\label{eq2}
E_0^{re/im} = \sum_m^{N_e} E_m^{re/im}
\end{equation}
where $E_0$ denotes the ground state energy and $E_m$ is the eigenvalue of $m$th state ($re/im$ corresponds to real or 
imaginary part). Since this is a non-Hermitian (NH) system, we observe both real and imaginary eigenvalues, necessitating 
the calculation of $E_0$ in both the real and imaginary domains. Hence, to compute the current in this case, we employ 
the following definition~\cite{pcnh1,pcnh2}.
\begin{equation}\label{eq3}
I^{re/im}_\phi = -c\frac{\partial{E_0^{re/im}}}{\partial\phi}
\end{equation}
where c is a constant. \par
For interacting case, we consider a interacting fermions with spin, and incorporate electron-electron interactions via an 
on-site interaction Hubbard term, $H_U$ in the Hamiltonian. The total Hamiltonian is expressed as
\begin{equation}\label{eq4}
H = H_0 + H_U
\end{equation}
where $H_0$ represents the non-interacting part of electrons with spin degrees of freedom together with incorporating the 
effects of AAH modulation and non-Hermiticity, while $H_U$ accounts for the on-site electron-electron interactions introduced 
by the Hubbard term. The non-interacting Hamiltonian considering spin degrees of freedom ($\sigma=\uparrow, \downarrow$) can 
be written as
\begin{equation}\label{eq5}
H_0 = t\sum_{j=1}^N \sum_\sigma (e^{i\theta}c_{j,\sigma}^\dagger c_{j+1,\sigma} + h.c.) + \sum_{j=1}^N \sum_\sigma \epsilon_j n_{j,\sigma}.
\end{equation}
Where $c_{j,\sigma}^\dagger (c_{j,\sigma})$ denote the fermionic creation (annihilation) operator and $n_{j,\sigma}$ represent the number 
operator for spin $\sigma(\uparrow,\downarrow)$ at site $j$.
The interacting part of the total Hamiltonian is given by
\begin{equation}\label{eq6}
H_U = U\sum_{j=1}^L n_{j,\uparrow}n_{j,\downarrow}
\end{equation}
where $n_{j,\uparrow} (n_{j,\downarrow})$ represent number operators for up (down) spin electrons. For small system size 
we calculate ground state energy by exact diagonalization (ED) and for larger systems ($\ge 20$ sites), we employ the 
Density Matrix Renormalization Group (DMRG)~\cite{dmrg1,dmrg2} method implemented in the ITensor\cite{dmrg3,dmrg4,dmrg5} 
library up to a bond dimension $\chi = 2000$. Since eigenstates in non-Hermitian matrices are nonorthogonal,  merely 
increasing the number of DMRG sweeps is not always sufficient\cite{dmrg6}. Hence, we employ the Arnoldi method, a more 
accurate eigensolver, to efficiently target the ground state and ensure reliable convergence in the presence of nonorthogonal 
eigenstates in non-Hermitian matrices\cite{dmrg7,dmrg8}. After calculating ground state energy, we use Eq.~\eqref{eq3} to calculate persistent current for interacting case.

\section{Numerical results and discussion}

In this section, we present numerical results based on the above theoretical prescription. The main focus is 
to discuss the intricate interplay between non-Hermiticity and disorder for both non-interacting and interacting 
cases. Before going to the detail analysis let us first specify the values of common parameters which are kept constant
throughout the calculations. The value of hopping parameter $t$ is set at $1$. Unless stated otherwise, we consider the 
system size $N=20$. Our analysis is conducted at half-filling (i.e, $N_e=N/2$) which offer an ideal framework for 
examining the system's behavior. All hopping integrals and energy eigenvalues are expressed in electron volts (eV).
The persistent current is measured in microamperes ($\mu$A). 

\subsection{Non-Interacting Case}
In this sub-section we discuss all the results in non-interacting picture i.e, the electron-electron correlation 
$U=0$. Here, the system is described by the TB Hamiltonian as written in Eq.~\eqref{eq1}, where 
we do not consider electron's spin degrees of freedom.

\subsubsection{Energy Spectrum and Current varying $\delta$}

\begin{figure}[ht]
	{\centering\resizebox*{8.5cm}{8.5cm}{\includegraphics{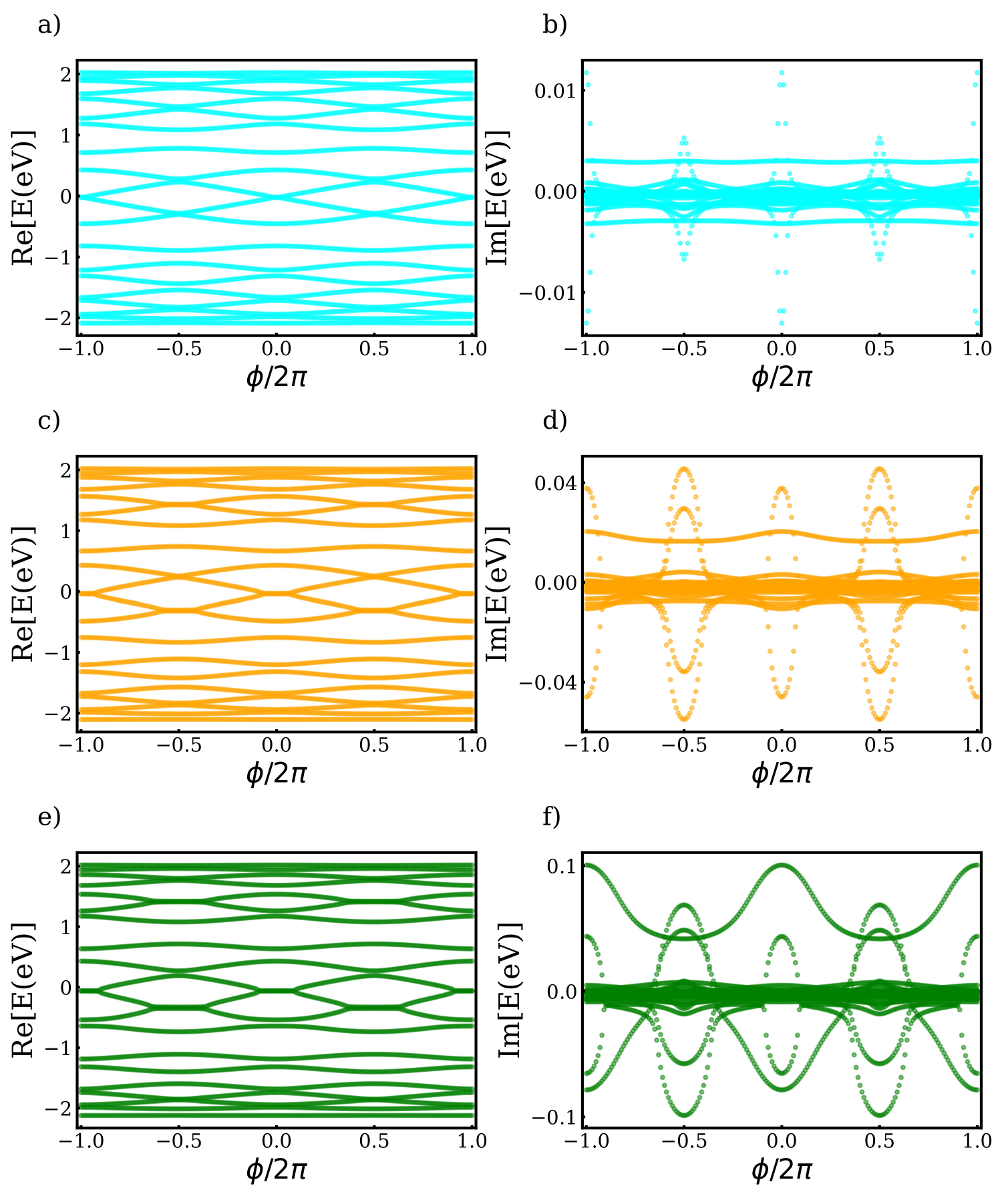}}}
	\caption{(Color online) Variation of real and imaginary energy eigenspectra as a function of AB flux for three 
		different values of $\delta$ when $W$ is set at $0.5$.  The cyan, orange, and green color spectra are for 
		$\delta=0.25$, $0.75$, and $1$ respectively.}
	\label{energyth}
\end{figure}

To understand the fundamental mechanism of persistent current, we begin by examining the energy eigenspectra of our 
system. In Fig.~\ref{energyth}, we plot the variation of energy spectrum as a function of flux. 
The left and right columns show the real and imaginary parts of the eigenvalues, respectively. 
Here we plot the energy spectrum for three different values of the complex phases, represented by three different 
colors. The cyan, orange, and green color curves denotes the energy spectra for $\delta=0.25, 0.75,$ and $1.0$ 
respectively. Here we set the disorder strength at $W=0.5$.
\begin{figure}[ht]
	{\centering\resizebox*{8.5cm}{6.5cm}{\includegraphics{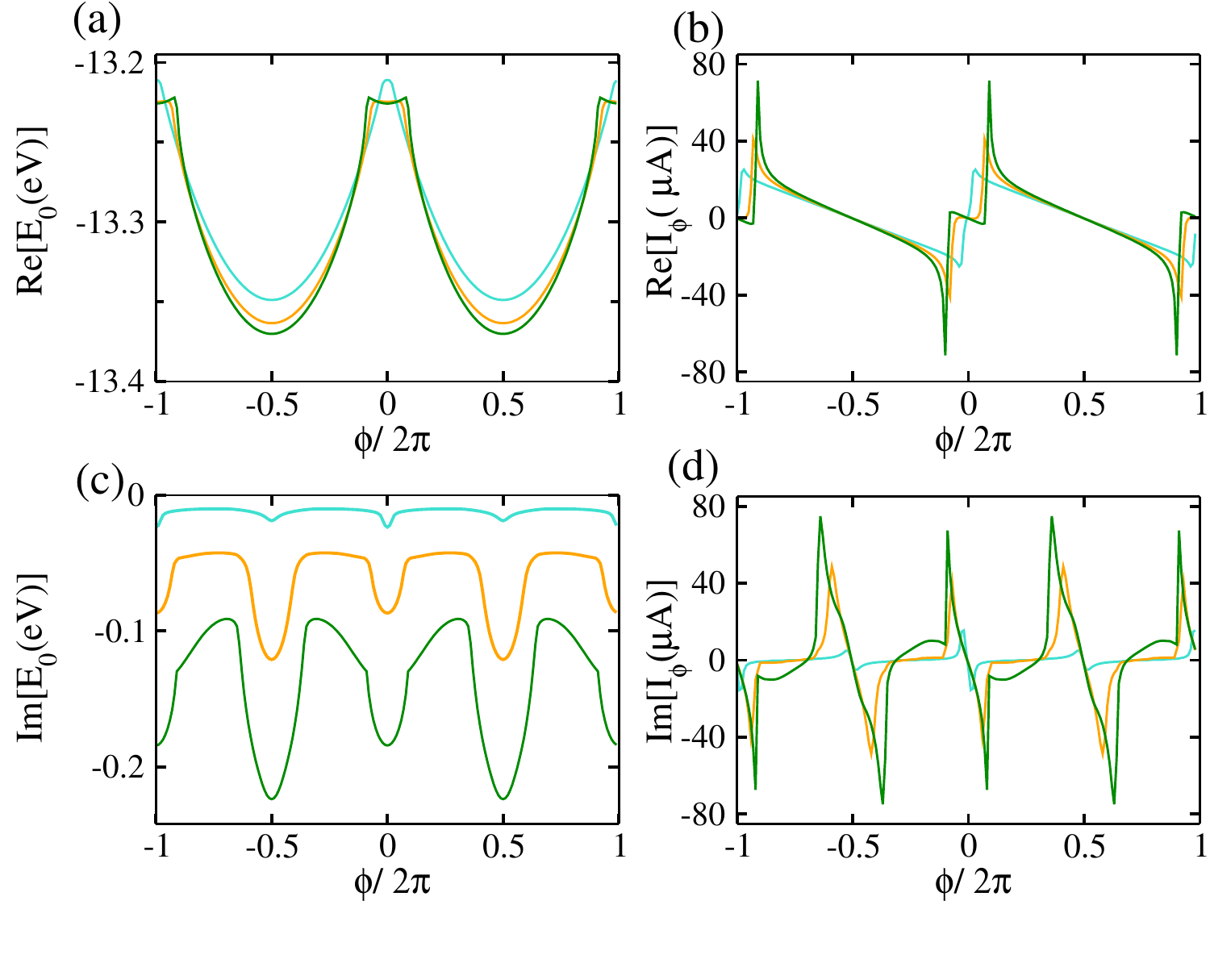}}}
	\caption{(Color online). Real and imaginary parts of the ground state energy versus AB flux ((a) and (c))and 
		that of  PC versus AB flux ((b) and (d))where cyan, orange and green curves represent the results for 
		$\delta=0.25, 0.75$ and $1$ respectively. Here we fix $W=0.5$.}
	\label{currenth}
\end{figure}
For $\delta=0.25$, the energy levels in the real spectrum (Fig.~\ref{energyth}(a)) have finite slopes and substantial 
degeneracy, some of these slopes get cancelled with each other. As a result current decreases in the system. In 
Fig.~\ref{energyth}(c) we can see that, for $\delta=0.5$, the degeneracy of the bands near the edges of the spectrum 
get lifted. With further increasing $\delta$, as shown in Fig.~\ref{energyth}(e), the degeneracy of most of the energy 
levels breaks. As a result fewer slopes cancel each other out, potentially enhancing the persistent current. In the 
imaginary part of the eigenvalues, in Fig.~\ref{energyth}(b), we observe that at $\delta=0.25$, most of the energy 
levels cluster around zero, forming a highly degenerate band. However, as the strength of the complex phase increases, 
as shown in Fig.~\ref{energyth}(d) and Fig.~\ref{energyth}(f), additional energy bands begin to emerge. Also the slopes 
of some energy levels increase with increasing $\delta$, implying a potential enhance in persistent current compared to 
$\delta=0.25$ case.

Now we examine how the ground state energy changes with magnetic flux, since the slope of the ground state energy is 
what essentially determines the PC. To calculate the real and imaginary ground state energies, the real and imaginary 
parts of the eigenvalues are sorted separately and added up to the filling factor. Since for our case we calculate for 
half filling condition, here we add the energies up to $N_e=N/2=10$ from the lowest energy level. In Fig. \ref{currenth}, 
we present a comparative visualization by plotting the ground state energy alongside the corresponding current for a 
clearer understanding. The upper row illustrates the real part of the ground state energy and the associated current, 
while the lower row displays the imaginary part of the ground state energy and its corresponding current, for three 
distinct values of $\delta$. Here we also take the values of $\delta$ as  $0.25, 0.75,$ and $1.0$ which are indicated 
by cyan, orange and green color respectively. The disorder strength is fixed at $W=0.5$. In Figs.~\ref{currenth}(a) 
and~\ref{currenth}(c), we can see that, both the real and imaginary parts of the ground state energy vary continuously 
with flux. However, at some points, the slope changes abruptly, commonly referred to as exceptional points. Both the 
real and imaginary parts of ground state energy depend on the value of $\delta$, but the imaginary part is more sensitive 
to the non-Hermiticity. The slopes of the ground state energy versus flux curve become steeper with 
increasing $\delta$, however for the imaginary part this change is more prominent.  

The dependence of ground state energy on $\delta$ is exactly reflected in the current flux characteristics. Due to 
the abrupt change in ground state energy at exceptional points, the current versus flux characteristics exhibit sharp 
spikes at these points. The real part of the current shown in Fig.~\ref{currenth}(b) increases with increasing the 
$\delta$ values. The enhancement is greater at the exceptional points. The imaginary part of the 
current shows more favorable response compared to the real current. As we can see in Fig.~\ref{currenth}(d), for 
$\delta=0.25$ current is much lower, but the current enhances significantly with increasing $\delta$ 
which is more prominent at exceptional points.
\subsubsection{Energy Spectrum and Current varying W}
\begin{figure}[ht]
	{\centering\resizebox*{8.5cm}{8.5cm}{\includegraphics{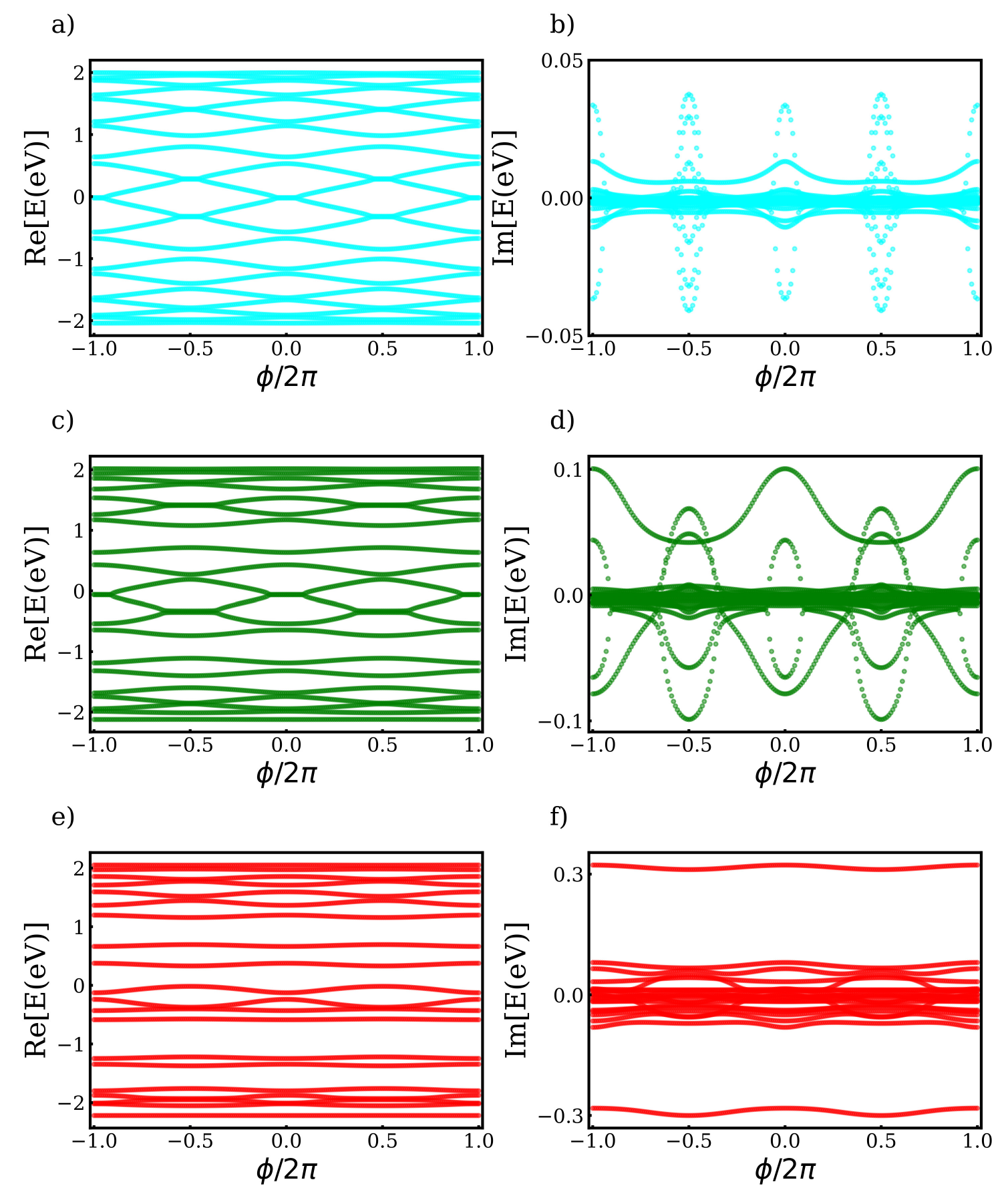}}}
	\caption{(Color online). Variation of real and imaginary energy eigenspectra as a function of AB flux for three 
	different values of $W$ with $\delta=1$. The cyan, green, and red  color curves are for $W = 0.25, 0.5$ and $0.75$ 
	respectively.}
	\label{energytw}
\end{figure}

Now we explore the effect of disorder strength on persistent current by fixing the non-Hermiticity factor $\delta=1$. 
Figure \ref{energytw} shows the variation of energy spectra with AB flux for different values of disorder strength. 
Cyan, green, and red curves correspond to $W=0.25, 0.5,$ and $0.75$ respectively. The left column shows spectra with 
real eigenvalues, while the right column shows those with imaginary eigenvalues. As we increase $W$ from  $0.25$ to 
$0.5$, the real part of eigenspectra shown in Figs.~\ref{energytw}(a) and \ref{energytw}(c), the degeneracy breaks for 
certain energy levels which in effect reduces the number of slope cancellations. This effect can potentially enhance 
the PC by allowing more energy levels to contribute constructively to the current flow. With the further increase in $W$, 
shown in Fig.~\ref{energytw}(e), most of the bands become flat which results decrease in PC again.
\begin{figure}[ht]
	{\centering\resizebox*{8.5cm}{6.5cm}{\includegraphics{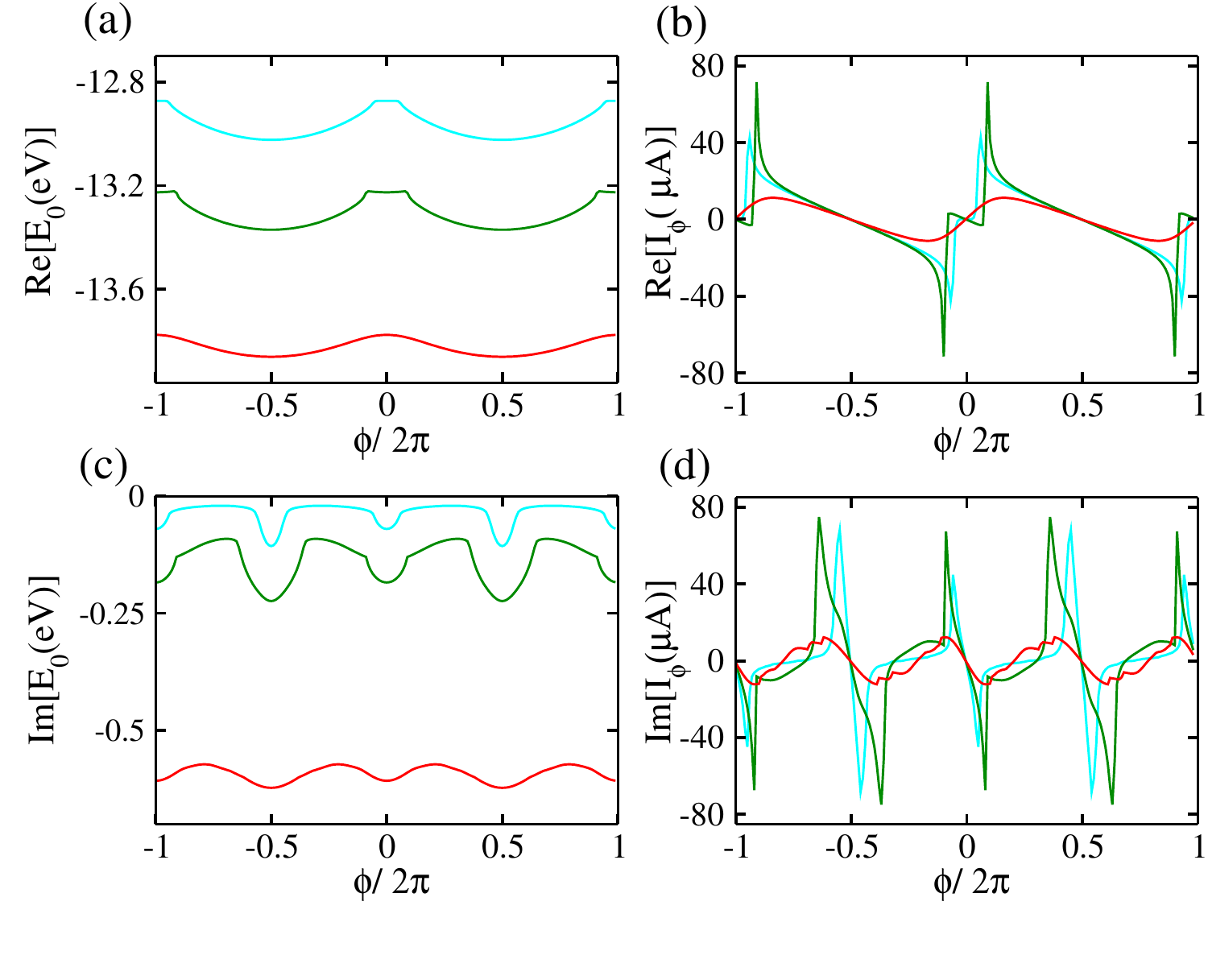}}}
	\caption{(Color online). Real and imaginary parts of the ground state energy versus AB flux ((a) and (c)) and real 
	and imaginary parts of PC versus AB flux  ((b) and (d))where cyan, green and red curves correspond to $W=0.25, 0.5$ 
	and $0.75$ respectively. Here we fix $\delta=1$.}
	\label{currentw}
\end{figure}

In the imaginary eigenspectra, we observe that at $W=0.25$ [Fig.~\ref{energytw}(b)], most energy levels cluster around zero, 
forming a highly degenerate band, except at a few exceptional points where certain energy levels spread out. As W increases 
to $0.5$ [Fig.~\ref{energytw}(d)], additional energy bands emerge, contributing to a rise in the persistent current. However, 
at $W=0.75$ as shown in Fig.~\ref{energytw}(f),  the spectrum develops a gapped nature with three distinct sub-bands. In the 
band near the middle of the spectrum, most slopes cancel each others out, leading to a reduction in the persistent current 
compared to the $W=0.5$ case.\par 

In Fig. \ref{currentw}, we illustrate the real and imaginary components of the ground state energies alongside their 
corresponding persistent currents for disorder strengths of $W=0.25, 0.5,$ and $0.75$. These are visually distinguished 
by cyan, green, and red colors, respectively. The upper and lower rows display the real and imaginary components of the 
ground state energy and current, respectively. Both the real and imaginary parts of the ground state energy exhibit 
continuous variation with flux, except at a few exceptional points. Notably, for $W=0.5$, the slopes of the ground state 
energy curve become steeper for both its real and imaginary components, with the latter experiencing a more pronounced 
change. As a result, the persistent current increases as the disorder strength increases from $W=0.25$ to $W=0.5$.
In this regime, the enhancement of the imaginary component of the current is more than that of its real counterpart. 
However, as W continues to increase, the spectrum becomes increasingly flattened in both components, ultimately resulting 
in a diminished PC.

\subsubsection{Interplay of Disorder and Non-Hermiticity}

From the above analysis it is found that, both the non-Hermiticity factor $\delta$ and disorder strength $W$ have an 
important role in PC. Here, we explicitly explore the effects of these two parameters on maximum current which is 
calculated by taking the maximum value of current in the entire flux range i.e, by varying $\phi$ from $0$ to $2\pi$. 
 \begin{figure}[ht]
 	{\centering\resizebox*{8.5cm}{4cm}{\includegraphics{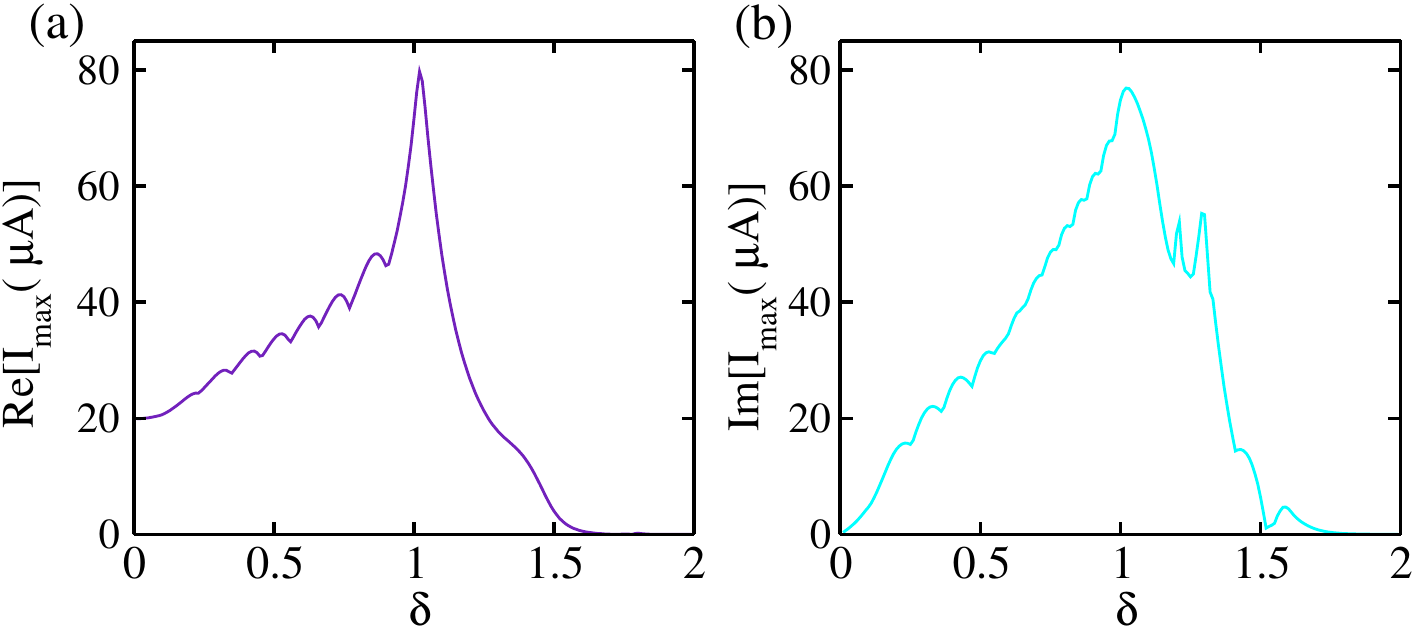}}
 	\resizebox*{8.5cm}{4cm}{\includegraphics{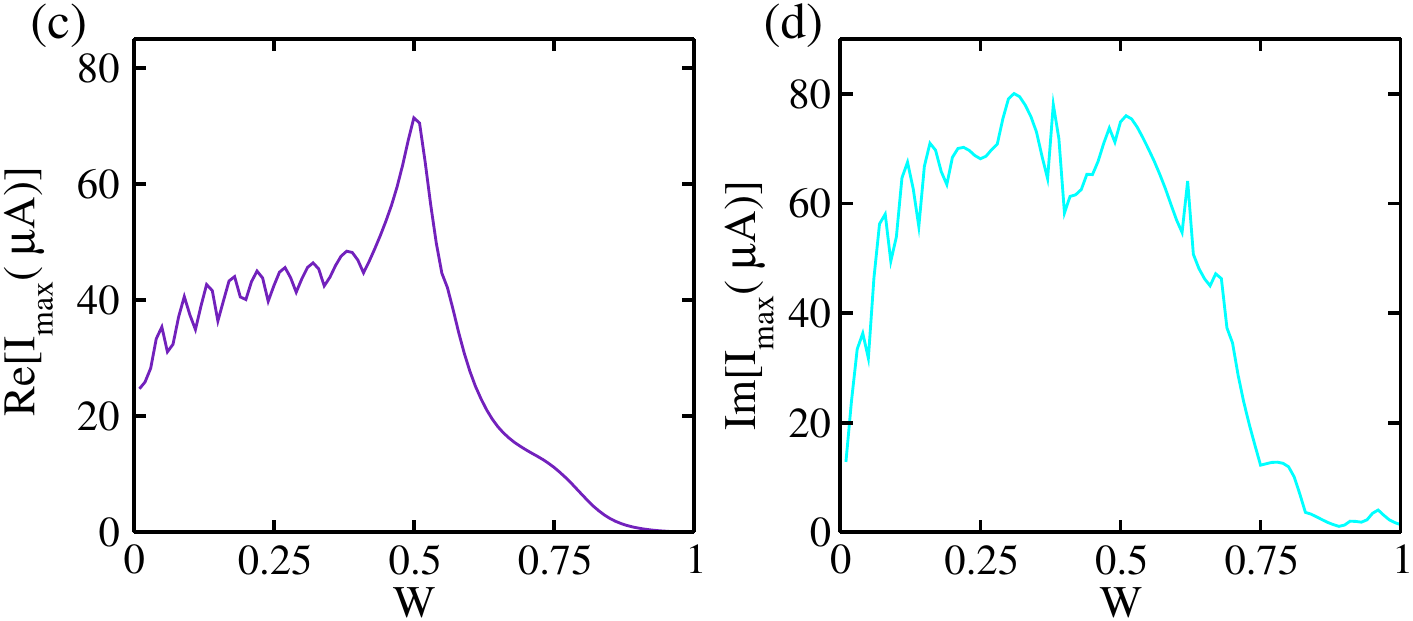}}}
 	\caption{(Color online). Upper panel: real and imaginary components of maximum current as a function of $\delta$ where 
 	$W$ fixed at $0.5$. Lower panel: response of real and imaginary components of maximum PC with disorder strength $W$, 
 	where $\delta$ fixed at $1$.}
 	\label{hh}
 \end{figure}
In order to further investigate the impact of the complex phase, $\delta$ and the modulation strength, $W$ 
of the AAH potential, we present in Fig. \ref{hh} the variation of the maximum persistent current as a function of both 
the parameters. Figures~\ref{hh}(a) and \ref{hh}(b) illustrates the dependence of the maximum persistent current on the 
complex phase $\delta$. As evidenced in Fig.~\ref{currentw}, the current exhibits an enhancement at $W=0.5$; thus, we 
fix this value to examine the variation of the persistent current with respect to $\delta$. The left panel showcases the 
maximum of the real component of the current, while the right panel illustrates the maximum of its imaginary counterpart. 
From Figs.~\ref{hh}(a) and \ref{hh}(b), it is evident that both the real and imaginary currents increase 
gradually with increasing $\delta$, reaching their peak at $\delta=1$. Beyond this point, as $\delta$ continues to increase, 
the current begins to decline. This result suggests that for a moderate degree of non-Hermiticity, the system can sustain 
an enhanced current even in the presence of disorder. Similarly Figs.~\ref{hh}(c) and \ref{hh}(d) show the 
variation of the maximum persistent current with AAH modulation strength $W$ in the presence of non-Hermiticity. Since  
at $\delta=1$, the maximum current is obtained, we fix this value here to calculate the results. Interestingly, we observe 
an unconventional behavior, the persistent current increases with disorder strength. This enhancement persist till $W=0.5$, 
after that current starts to decline (more details regarding the response of maximum current at different parameter values is provided in the Supplemental Material\cite{sm}). This behavior arises from the specific nature of the AAH modulation. 
The imaginary phase factor introduces both real and imaginary components into the on-site potential. The imaginary part 
effectively represents site-dependent loss and gain within the system. The imaginary component becomes positive at some 
sites and negative at others and as a result the system behaves as a disordered ring with spatially varying gain and loss.
The strength of this gain/loss imbalance depends on both the parameters, $\delta$ and $W$. For small values of both parameters, the system 
approximates a balanced gain-loss configuration. As these parameters increase, the imbalance becomes more pronounced. In 
particular, at sites where the potential is positive, the electron's wavefunction acquires a phase that enhances the 
probability of the electron being localized at those sites, whereas at sites with negative potential lead to a reduction of 
the probability for being at that site~\cite{lossgain}. In our system, the magnitude of the negative imaginary components 
tends to be larger than that of the positive ones with increasing $\delta$ and $W$. This asymmetry discourages electrons 
from remaining at certain lattice sites, effectively promoting delocalization and enhancing the current. However, with 
further increase in both $\delta$ and $W$ enhances the disorder in the system, which in turn suppresses the current. 
This non-trivial trend implies that in presence of non-Hermiticity, moderate disorder strength can enhance persistent 
current.

\begin{figure}[ht]
	{\centering\resizebox*{8.5cm}{4.5cm}{\includegraphics{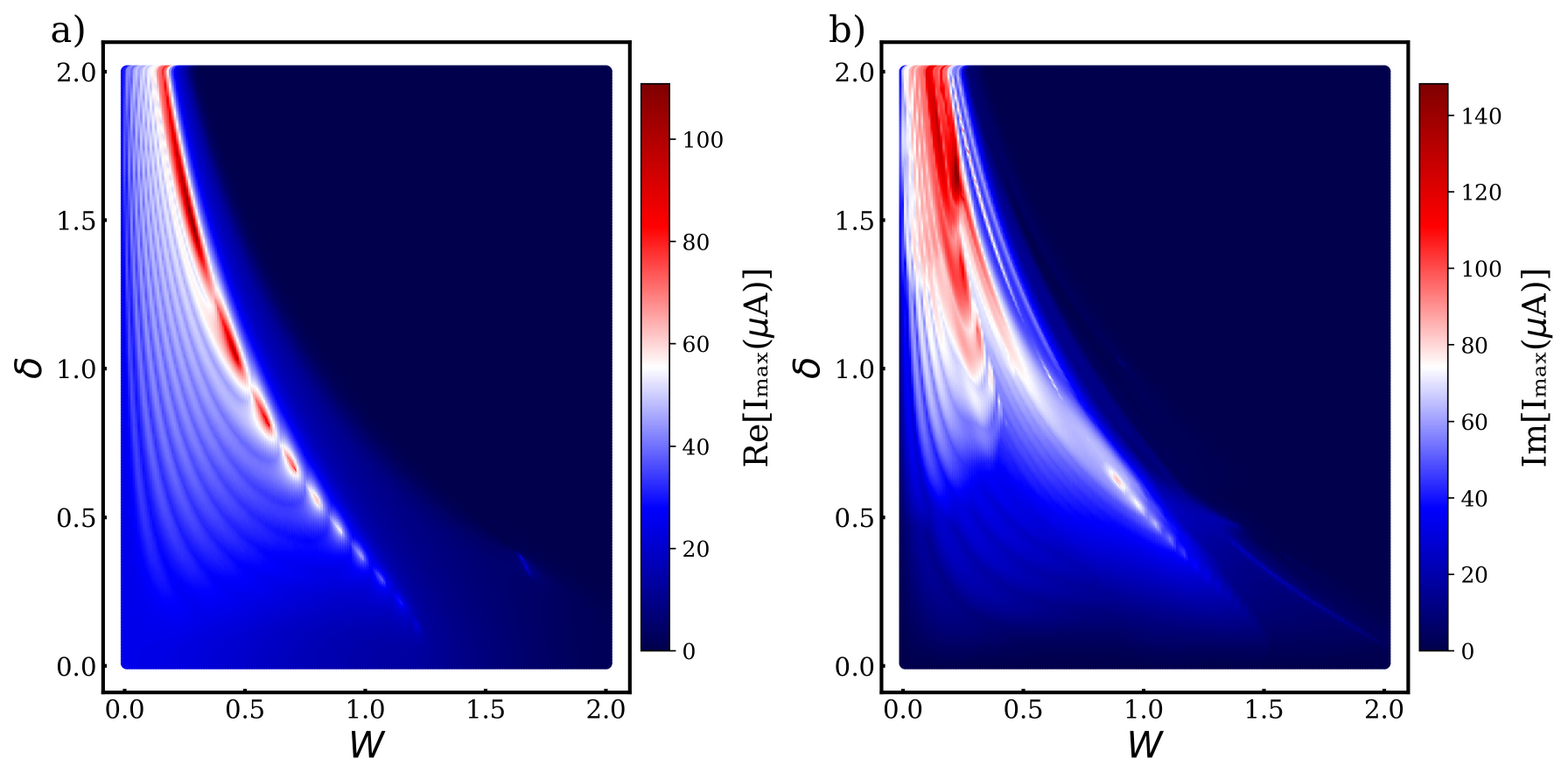}}}
	\caption{(Color online). Simultaneous variation of real and imaginary parts of maximum PC with $\delta$ and $W$.}
	\label{density}
\end{figure}

Figure~\ref{density} shows the simultaneous variation of maximum persistent current with disorder strength and non-Hermiticity. 
This visualization allows us to identify the specific regions in parameter space where
the maximum persistent current reaches its peak by changing $\delta$ and $W$, providing deeper insight into the interplay 
between AAH modulation strength and complex phase in governing the system's transport properties. Here we can see that, 
for a wide range of parameter values PC is reasonably high. The maximum persistent current occurs in the regime where the 
disorder strength, W is moderate, while the complex phase, h is relatively higher. This suggests that non-Hermiticity can 
enhance transport with increasing disorder value whereas, as expected, excessive disorder suppresses it, highlighting the 
intricate interplay between quasiperiodic modulation and non-Hermiticity in determining the system’s conductive properties.

\subsubsection{Variation with system size}
\begin{figure}[ht]
	{\centering\resizebox*{4cm}{3.3cm}{\includegraphics{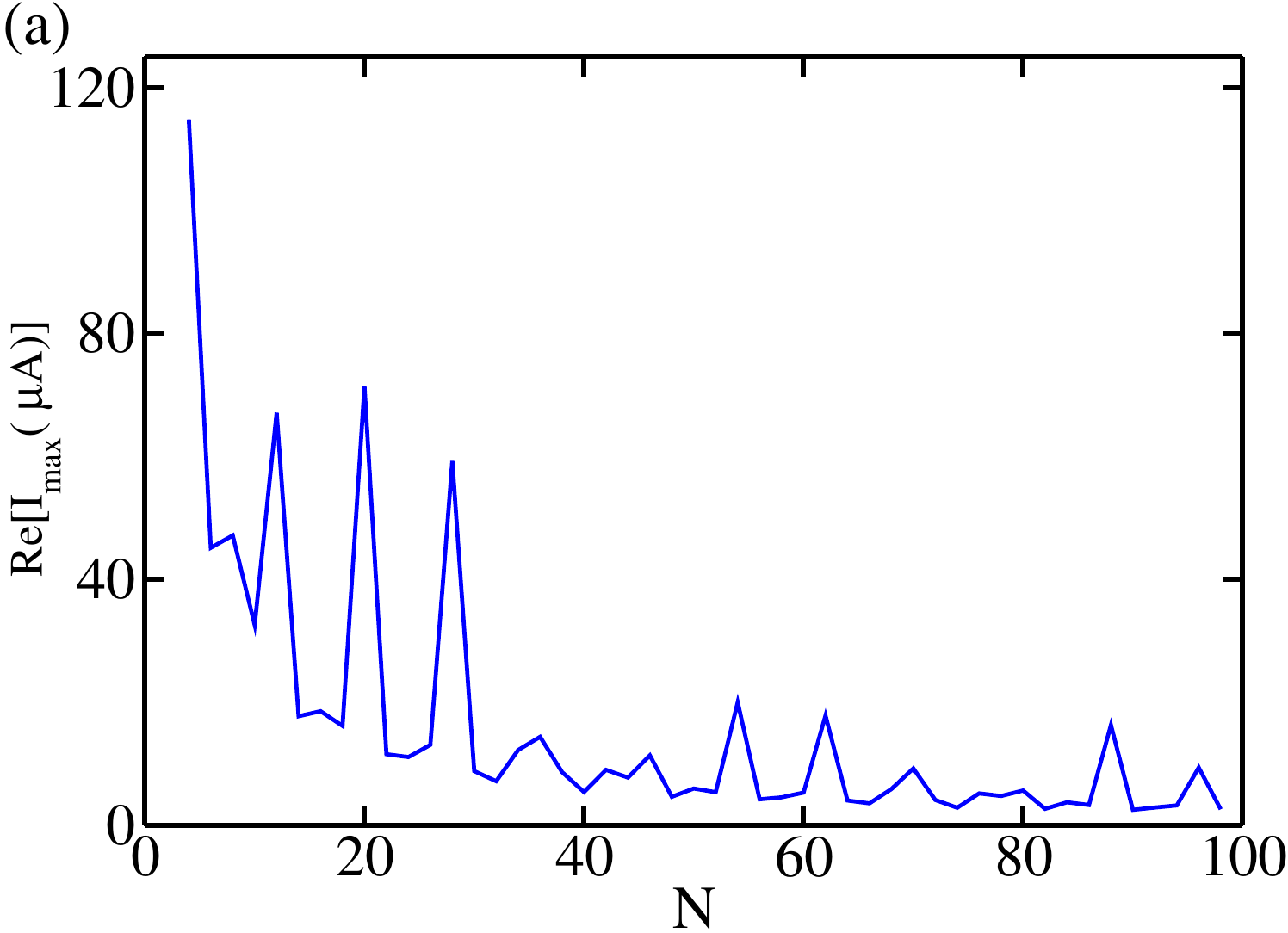}}
		\resizebox*{4cm}{3.3cm}{\includegraphics{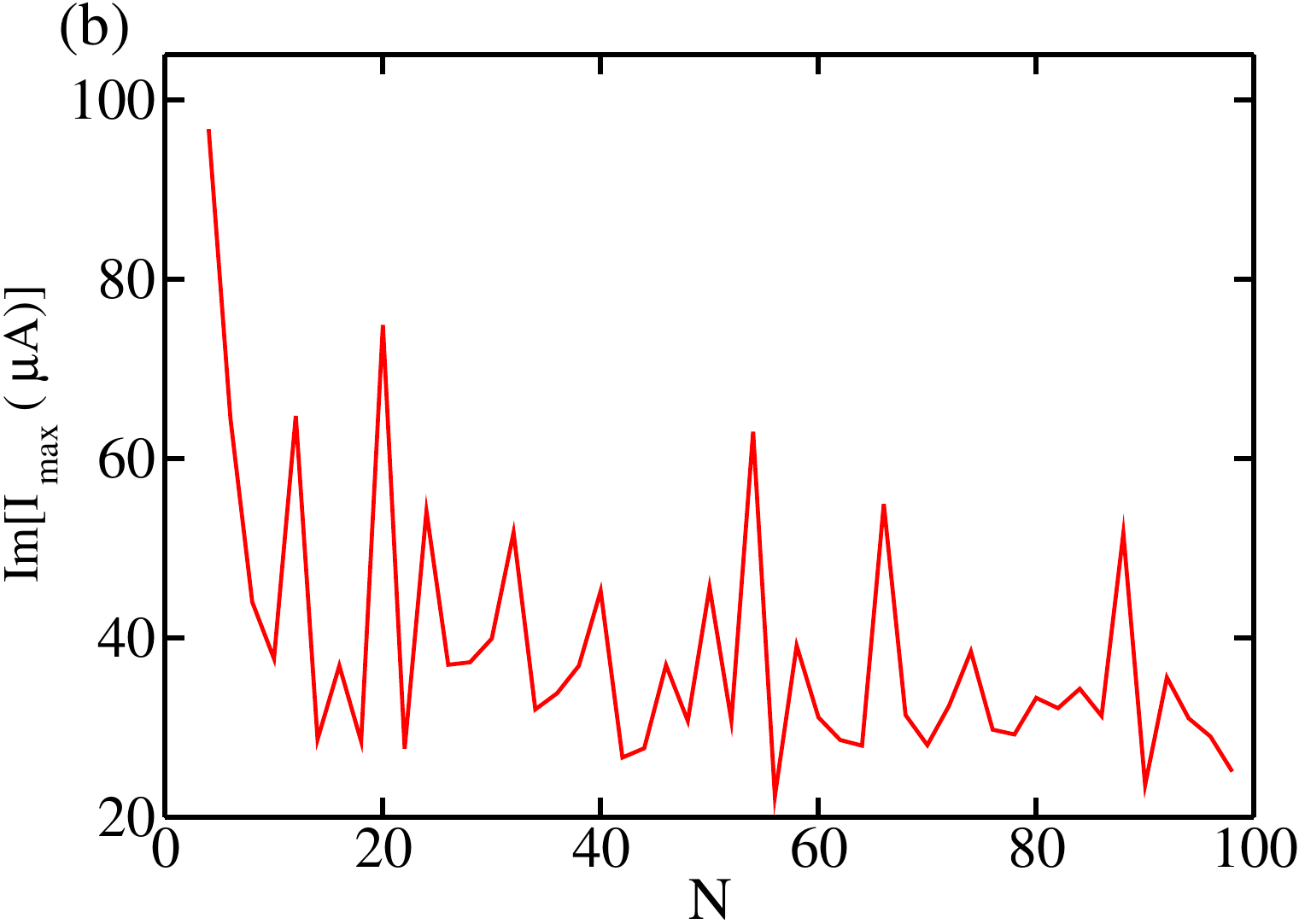}}\par}
	\caption{(Color online). (a) Real and (b) imaginary parts of maximum current as a function of system size. 
	Other parameters are fixed as: $\delta=1$, $W=0.5$.}
	\label{ss}
\end{figure}

In Fig.~\ref{ss}, we show the variation of persistent current with system size. We calculate the results for half-filled 
condition, to ensure this we vary the system size by taking the interval $\Delta N=2$, so that we have even number of sites 
in the systems. The left figure represents the variation for real current whereas right one indicates that for imaginary 
current. Here we fix the non-Hermiticity and disorder strength at $1$ and $0.5$ respectively.
For both the real and imaginary currents, fluctuations arise as a consequence of quantum interference effects. In addition 
to these fluctuations, the magnitude of the current diminishes with increasing system size in both cases, reflecting the 
typical behavior of persistent currents in larger systems. However, a moderate magnitude of the imaginary current persists 
even at higher system sizes. This is due to fact that the imaginary current arises for the imaginary phase 
of the AAH modulation and we have seen in our previous discussions that the non-Hermiticity helps to enhance the current in 
presence of moderate disorder. This enhancement is more prominent for the imaginary current which helps to maintain a moderate 
current even at higher system sizes.

\subsection{Interacting Case}
In this subsection we present the results for interacting case considering electron's spin. The system is described by 
the Hubbard Hamiltonian as represented by Eq.~\eqref{eq4}. Here we explore the effect of Hubbard correlation 
$U$ on PC in presence of non-Hermiticity and disorder. 

 \begin{figure}[ht]
	{\centering\resizebox*{8.5cm}{4cm}{\includegraphics{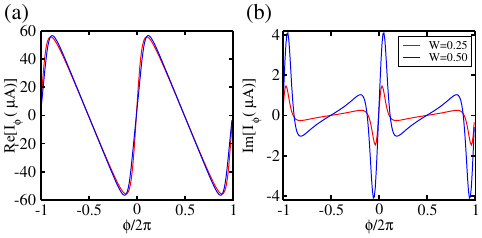}}
		\resizebox*{8.5cm}{4cm}{\includegraphics{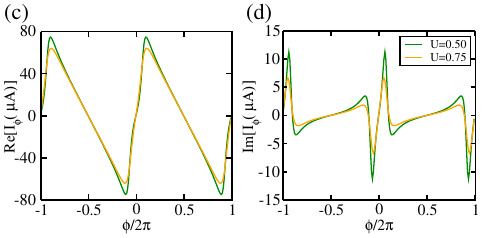}}}
	\caption{(Color online). Upper panel: Real and imaginary components of persistent current as a function 
			of $\phi$ in the presence of onsite Hubbard interaction $U = 1$ for two different values of $W$. Red color 
			correspond to $W = 0.25$ and blue color represents $W = 0.5$. Lower panel: Real and imaginary components of 
			persistent current as a function of $\phi$ in the presence of disorder $W = 0.5$ for two different values of 
			$U$. Green color correspond to $U = 0.5$ and orange color represents $U = 0.75$. Here we choose $\delta = 1$ 
			and $N = 12$.}
	\label{int1}
\end{figure}

Now to investigate whether the observed trends in the non-interacting case hold in the presence of local many-body effects, 
we consider a finite size ring consist of 12 sites described by the Hubbard Hamiltonian with non-Hermitian AAH term. 
We apply an exact diagonalization (ED) routine to the matrix to obtain both the real and imaginary eigenvalues along with 
their corresponding eigenvectors.
\begin{figure}[ht]
	{\centering\resizebox*{8.5cm}{4.5cm}{\includegraphics{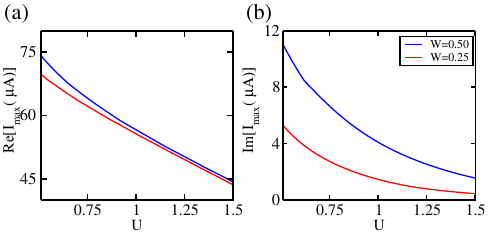}}}
	\caption{(Color online). (a) Real and (b) imaginary parts of maximum current as a function of the 
	onsite Hubbard interaction strength $U$ for two different values of $W$ with $\delta=1$ and $N=12$}
	\label{int3}
\end{figure}
Figures~\ref{int1}(a) and \ref{int1}(b) show the real and imaginary part of the persistent current
as a function of flux for two different values of disorder strengths respectively. The PCs in 
this instance are represented in blue color for a disorder strength of $W=0.25$ and in red color for $W=0.5$ where we fix 
$\delta=1$. In the real part, we observe only a slight enhancement in PC with increasing disorder strength, whereas the 
imaginary part exhibits a more pronounced increase, indicating a stronger influence of non-Hermiticity on the imaginary 
component of the current. Notably, even in presence of electron interactions ($U=1$), we observe an increase in PC as 
disorder strength rises from $W = 0.25$ to $W = 0.5$.

Figures~\ref{int1}(c) and \ref{int1}(d) present the real and imaginary components of the PC for two 
different values of the local on-site Hubbard interaction strength, $U$. In both the components, a clear reduction in the 
current magnitude is observed with increase in $U$ values. To further illustrate, in Fig.~\ref{int3}, we plot  the 
maximum amplitude of the persistent current as a function of $U$ for two different values of $W$. Here we can see that 
the enhancement of the current magnitude occurs with increasing disorder strength as previously observed in Figs.~\ref{int1}(a) 
and \ref{int1}(b), which persists across various onsite Hubbard interaction strengths. However, at larger values of 
$U$, the current decreases sharply. These behaviors can be attributed to the half-filled case, where 
strong on-site repulsion favors configurations with single occupancy per site. As $U$ increases, this constraint becomes 
more pronounced and the electron mobility reduces significantly. The repulsive interaction inhibits hopping processes 
due to the energetic cost of double occupancy. Consequently, for sufficiently large $U$, the current diminishes and 
eventually vanishes as nearly all sites become singly occupied, effectively freezing the dynamics.

The overall trends observed in the non-interacting case remain consistent even in the presence of the local on-site 
Hubbard interaction, indicating that the fundamental effects of non-Hermiticity and quasiperiodic modulation persist 
despite electron-electron interactions.

\begin{figure}[ht]
	{\centering\resizebox*{8.5cm}{4.5cm}{\includegraphics{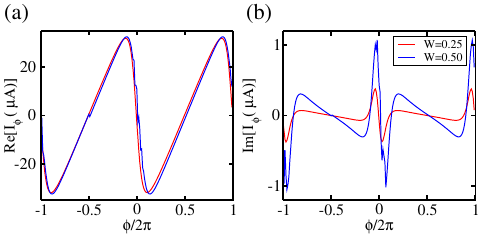}}}
	\caption{(Color online). Real and imaginary components of persistent current as a function of $\phi$ in the 
		presence of onsite Hubbard interaction $U$ for two different values of $W$ for system size $N=20$. Red color 
		correspond to $W=0.25$ and blue color represents $W=0.5$. Here we choose $U=1$ and $\delta=1$.}
	\label{int2}
\end{figure}
To further validate our findings, we examine whether this effects persist in larger system sizes. We employ the DMRG 
technique to compute both the real and imaginary components of the ground state energy. Figure~\ref{int2} illustrates 
the modulation of both the components of the PC for a system with N = 20 sites as a function of phase, considering two 
different disorder strengths with non-Hermiticity factor $\delta=1$. Similar to our ED studies, we observe only a slight 
enhancement in the real part of the PC. However, as expected, the imaginary component of the PC exhibits a more pronounced 
increase as the disorder strength increases. Hence, our findings for larger systems with electron interactions further 
reinforce the interplay between non-Hermiticity and quasiperiodic disorder in enhancing transport properties.
\vskip 0.1in

\noindent \textbf{Experimental realization:}
Now, for the sake of completeness, we would like to highlight the potential experimental realizations of the system proposed 
in this work. Quantum rings have been successfully fabricated in various experimental platforms, with persistent currents 
unequivocally observed in such systems~\cite{exp1,exp2}. However, our model consists of a ring with complex onsite 
Aubry-Andr\'{e}-Harper (AAH) potential, wherein non-Hermiticity arises through a complex phase in the potential. To realize 
the electronic analogue of our theoretical framework, one requires a tunable mesoscopic quantum ring wherein the quasiperiodic 
disorder can be precisely engineered~\cite{exp3}. Such a system with imaginary phase factor in quasiperiodic potential,  
has already been implemented for localization studies using  non-Hermitian photonic quasicrystals~\cite{exp4,exp5}.
For implementing our model, atomic sites may be spatially arranged in a ring geometry via molecular beam epitaxy (MBE) 
techniques~\cite{exp6,exp7}, or alternatively, through ultracold atomic lattice platforms, such as a Bose-Einstein condensate (BEC) 
of $^{39}$K atoms~\cite{exp8,exp9}. The quasiperiodic potentials considered in the model may be implemented by a controlled 
application of a series of standing wave laser beams. In our proposed setup, the real part of the persistent current can 
be extracted from the system's magnetization, whereas the imaginary component, arising from non-Hermitian effects can be 
inferred by monitoring the time derivative of the magnetization~\cite{exp3}.

\section{summary and outlook}
In summary, this study has explored the impact of disorder and non-Hermiticity on the persistent current in 
a mesoscopic ring subjected to an AB flux. The disorder and non-Hermiticity have been introduced via the AAH 
model, with the latter arising from the phase of the on-site AAH modulation. We have analyzed the results for 
non-interacting as well as interacting cases. For non-interacting scenario, the system has been described 
using a TB Hamiltonian for spinless fermions, and the ground state energy has been obtained through diagonalization.
Whereas for interacting case, we have incorporated spin degrees of freedom and included repulsive on-site Hubbard 
interaction. The ground-state energy has been computed using exact diagonalization for smaller system sizes, while 
for larger systems, the DMRG method has been employed. By differentiating 
the ground state energy with respect to the AB flux, we have computed the persistent current and have observed 
both its real and imaginary components due to non-Hermiticity. We have thoroughly analyzed the influence of 
various system parameters on these components of PCs. 
All calculations have been performed for a half-filled system. The key findings of our study have 
been as follows.\\
$\bullet$ The interplay between disorder strength($W$) and non-Hermiticity ($\delta$) is very promising. For non-zero 
disorder strength, both the real and imaginary parts of the PC increases with $\delta$ and reaches its maximum value 
at sufficiently high value of $\delta$.\\
$\bullet$ The systematic study of the interplay between the AAH strength and non-Hermiticity reveals that, in 
presence of $\delta$, an enhancement of PC occurs with disorder strength and it increases till a critical value of $W$.\\
$\bullet$ The simultaneous variation of $W$ and $\delta$ indicates that the enhancement of PC takes place for a wide 
range of parameter values.\\
$\bullet$ Numerical simulations indicate a behavior similar to the non-interacting case for moderate interaction 
strength, suggesting that the interaction effect does not significantly alter the characteristics of PC.\\
$\bullet$ The DMRG results indicate that, in the presence of Hubbard interaction, the persistent 
current exhibits an enhancement with increasing disorder strength in both its real and imaginary components, although, 
this enhancement is more pronounced in the imaginary component compared to the real one.\\ 
$\bullet$ Notably, the enhancement in PC has been attained without relying on any hopping dimerization mechanism
which proves that our results are applicable to any non-Hermitian quasiperiodic ring without imposing any constraints 
in the system.\\
Our results highlight a potential regime where non-Hermitian quasiperiodicity even in presence of electron-electron correlations 
can enhance persistent currents, offering new insights in the future research of non-Hermitian
quasiperiodic systems. 

\section*{ACKNOWLEDGMENTS}

SS is thankful to DST-SERB, India (File number: PDF/2023/000319) for providing her research fellowship. SS acknowledges JNCASR, India for funding.
SKP acknowledges the JC Bose fellowship and SERB, Govt. of India for the financial assistance.

\end{document}